# Differentiation between shallow and deep charge trap states on single poly(3-hexylthiophene) chains through fluorescence photon statistics

Kristin S. Grußmayer, Florian Steiner, John M. Lupton, Dirk-Peter Herten, and Jan Vogelsang

**Abstract:** Blinking of the photoluminescence (PL) emitted from individual conjugated polymer chains is one of the central observations made by single-molecule spectroscopy (SMS). Important information, e.g., regarding excitation energy transfer, can be extracted by evaluating dynamic quenching. However, the nature of trap states, which are responsible for PL quenching, often remains obscured. We present a detailed investigation of the photon statistics of single poly(3-hexylthiophene) (P3HT) chains obtained by SMS. The photon statistics provide a measure of the number and brightness of independently emitting areas on a single chain. These observables can be followed during blinking. A decrease in PL intensity is shown to be correlated with either (i) a decrease in the average brightness of the emitting sites; or (ii) a decrease in the number of emitting regions. We attribute these phenomena to the formation of (i) shallow charge traps, which can weakly affect all emitting areas of a single chain at once; and (ii) deep traps, which have a strong effect on small regions within the single chains.

Single-molecule spectroscopy (SMS) is the only technique that yields information regarding dynamic fluctuations in photoluminescence (PL) without the need for synchronization between molecules.[1, 2] It is therefore not surprising that SMS of conjugated polymers (CPs) exposes the highly dynamic changes in PL intensity that occur on a single chain.[3] Single-step PL blinking events have been observed and attributed to efficient intramolecular excitation energy transfer to a PL-quenching polymer defect, which is thought to be a positive charge on a polymer chain.[3-6] This positive charge resides in a trap state and quenches successively formed excitons on the chain,[7] which are funneled towards the quenching site with varying efficiency. The depth of a quenching event is a measure of the amount of material affected by the quencher,[8] which is strongly morphology dependent and can even affect multiple chains simultaneously.[9-12] The implications of this quenching for light-emitting diodes and sensors are apparent, and this effect has consequently been studied in great detail.[3] For example, the diffusion lengths of excitons can be extracted by combining the observation of PL quenching events with super-resolution fluorescence microscopy techniques.[13-15]

However, it is not possible to extract information on the nature of electronic trap states holding the positive charge by merely evaluating PL quenching events through the brightness of the molecule. It has been suggested that there are at least two main types of trap states in CPs: (i) shallow trap states,[16-18] in which the charge can be highly mobile and can therefore access a large amount of material, and (ii) deep trap states, in which the charge is immobile and confined within a small region of the material.[16-20] Fortunately, in principle, the photon statistics obtained from a single CP chain contain the information required to differentiate between quenching events arising from either shallow or deep traps, as will be outlined in the following.

Photon statistics are typically used to count the number of *independently* emitting entities – fluorophores – present within the observation volume of a confocal fluorescence microscope.[21-24] Such counting is achieved by exploiting the fact that a single fluorophore can only emit one photon, two fluorophores up to two photons and so forth per excited state lifetime, i.e., the PL lifetime. Of course, this simple picture is not applicable to CPs because the fluorophores on a single chain – the chromophores, over which π-conjugation extends – are *connected* to each other.[3] Singlet-singlet and singlet-triplet annihilation due to excitation energy transfer leads to single-photon emission – photon antibunching – from multiple fluorophores and pronounced PL quenching, respectively.[5, 25] Therefore, the photon statistics of a single CP chain cannot be used to quantify the number of fluorophores present, but it can be used to count the average number of effectively independent emissive regions in a single chain in which efficient singlet-singlet annihilation takes place. The conjugated segments of a CP, the chromophores, can only be differentiated by their spectrum,[26-30] not by photon statistics.

Here, we demonstrate how to monitor changes in photon statistics during PL-quenching events on single poly(3-hexylthiophene) (P3HT) chains, a device-relevant material used in organic solar cells (structure shown in Scheme 1a).[31] To reliably assess the photon statistics and enable counting of the number of emitters within a single weakly fluorescent P3HT chain, an advanced Hanbury Brown-Twiss geometry with four detectors and a sophisticated analysis method was employed.[32-34] Two different types of quenching events were observed, which we attribute to shallow and deep charge traps.

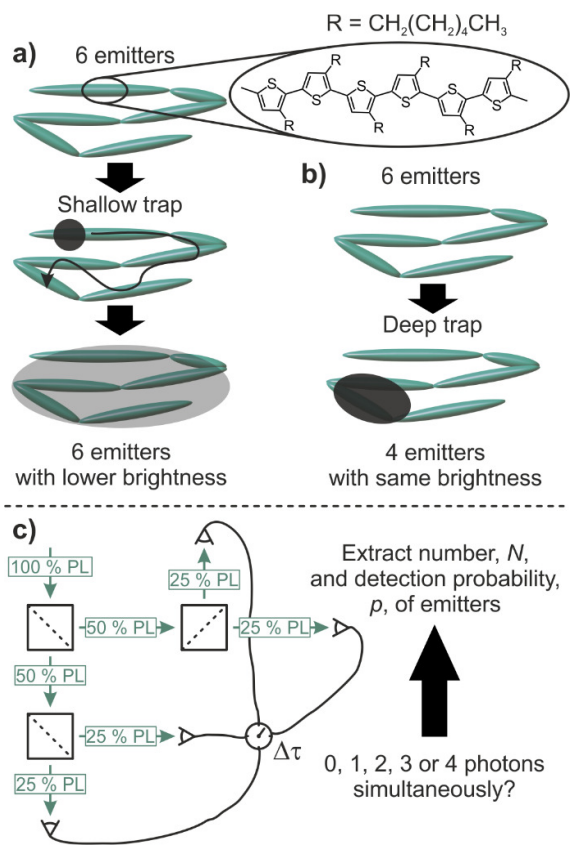

**Scheme 1.** *Effect on fluorescence photon statistics of the formation of shallow and deep charge carrier trap states in conjugated polymers. Poly(3-hexylthiophene) is used as a model system, the structure of which is shown in (a). (a, b) An independent emissive region (green ellipse) on a conjugated polymer chain is defined as a region of the*

*molecule that behaves as a single quantum emitter. Efficient singlet-singlet annihilation occurs within this region since energy is funnelled to the lowest-lying electronic state, an exciton trap. (a) Example of a conjugated polymer consisting of 6 emitters on which a shallow charge trap state is formed. The shallow trap affects all emitters approximately equally since the charge is virtually free to move. This results in an overall decrease in the photoluminescence (PL) of the entire polymer chain due to quenching without changing the number of emitters: in this case, the degree of photon anti-bunching stays the same. (b) Example of a conjugated polymer with formation of a deep charge trap state. A small number of emitters is affected by the quencher. However, in this case the effect of the quencher on the individual specific chromophore is stronger than in the case illustrated in (a). Quenching again results in a decrease in PL but additionally, the number of emitters is now also lowered, implying that the degree of photon anti-bunching must change. The working principle of counting by photon statistics (CoPS) is demonstrated in (c). The overall PL is detected by four equally weighted detectors, which are connected to a time-correlated single photon counting (TCSPC) card. The number, N, and detection probability, p, of the emitters is extracted from the amount of detecting 0, 1, 2, 3 or 4 photons simultaneously and comparing this to a mathematical model as described in the supporting information.*

Scheme 1 illustrates the connection between shallow and deep trap formation and the different expected effects on the photon statistics. Panels a and b explain what constitutes a single emitter in the context of photon statistics (green ellipse); a single emitter is defined as a region within the polymer chain in which efficient singlet-singlet annihilation occurs. Different chromophores on the chain will have different energies,[10] so that an energy gradient exists over a certain region, driving the excitation to the lowest energy state. Such an arrangement of multiple chromophores within a single chain is referred to as an energy funnel.[35, 36] A single such energy funnel will behave as a single emitter in terms of its photon statistics, i.e., the funnel shows photon antibunching since all chromophores within the funnel undergo singlet-singlet annihilation. Under the assumption that these energy funnels are all of the same size, the photon statistics will directly report on the number of coupled chromophore regions that exist and the brightness of each such region. The number of emissive regions within the chain will be determined by the exciton mobility: increasing mobility will raise the number of chromophores coupled, lowering the overall number of emitters, which are capable of emitting photons simultaneously. However, the mobility of charge exciton quenchers is also crucial. The lower the mobility of such carriers, i.e. the stronger carrier trapping, the more selective the charge quenching will be with respect to the molecular emitters. Extracting single-emitter brightness *p* and the number of emitters *N* from photon statistics will therefore provide information on the nature of the quenching species. Panel (a) illustrates the case of a CP chain containing 6 independent emitters (energy funnels) and the case of a shallow trap. The charge responsible for PL quenching is capable of screening the complete CP chain and affects all emitters simultaneously, assuming the duration of this process is shorter than the average PL lifetime. As a result, given such global quenching the degree of photon anti-bunching will not change during a partial PL quenching event. The number of independent emitters remains the same, but the average brightness is reduced. This result stands in contrast to that in the case of a deep trap, shown in panel (b): Here, only a few emitters are affected by a quencher, but more strongly than in the previous case, even up to a point at which no photons are emitted. In this case, the degree of photon anti-bunching is altered because fewer emitters are active.

However, the average brightness per emitter persists because the remaining emitters are not affected by the quencher.

Dynamically monitoring the photon statistics of a single CP chain with low PL quantum yields is already challenging, but especially so during the partial quenching of PL. The standard procedure involves the 50/50 splitting of the PL between two detection channels and the calculation of the second-order cross-correlation function $g^{(2)}(\Delta\tau)$, where $\Delta\tau$ is the time delay between two detection events on both detectors. The ratio of $g^{(2)}(\Delta\tau = 0)$ to $g^{(2)}(\Delta\tau \gg \tau_{PL})$, with $\tau_{PL}$ the PL lifetime, is directly related to the number, $N$, of emitters: $g^{(2)}(\Delta\tau = 0)/g^{(2)}(\Delta\tau \gg \tau_{PL}) = 1 - 1/N$.[32] With this method, only a small number (up to three) of independent emitters can be resolved, which is not sufficient for evaluating the photon statistics of single CP chains.[32] A recently introduced method to count the number of independent emitters through fluorescence correlation, referred to as "counting by photon statistics" (CoPS), is based on four equally weighted detectors and fitting the measured distribution of detecting zero, one, two, three or four photons simultaneously per excitation cycle to a theoretical model and yields a much higher precision and an extension of the counting range for the number of emitters at very low PL intensities.[34] The basic idea is sketched in Scheme 1c. Details of the working principle are given in the Supporting Information and in ref. [32-34]. In short, two observables are extracted by this method, first the number of independent fluorophores, $N$, and second the detection probability, $p$, defined as $p = I_{laser}\sigma_{abs}\Phi_f g/(f_{rep}h\nu)$, with $g$ accounting for the collection efficiency of the microscope, $\sigma_{abs}$ and $\Phi_f$ describing the absorption cross-section and fluorescence quantum yield of the fluorophore, respectively, and $I_{laser}$ corresponding to the excitation intensity in the focal volume divided by the photon energy $h\nu$ and the laser repetition frequency $f_{rep}$. Therefore, $p$ is a measure for the average molecular brightness per independent emitting species.

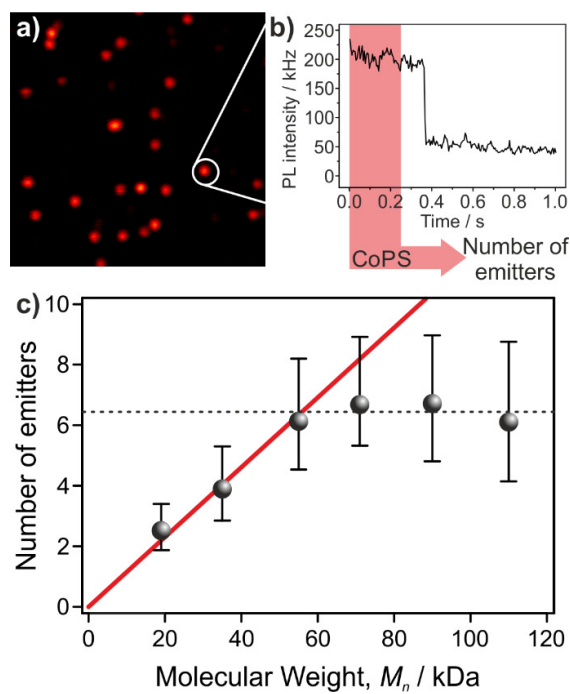

*Figure 1. Relationship between number of emitters, N, and molecular weight of poly(3-hexylthiophene) (P3HT). (a) Scanning confocal fluorescence image (20×20 μm²) of single P3HT chains embedded in Zeonex excited at 470*

*nm with an excitation intensity of 4 kW/cm². (b) One P3HT chain at a time was placed inside the focus to record PL transients. Only the first 250 ms of each transient was used (red shaded area) to obtain the number of emitters by employing the counting by photon statistics (CoPS) algorithm. (c) Number of emitters versus the number average molecular weight, $M_n$, of the P3HT sample used. The median of number of emitters is displayed along with the lower and upper quartiles as error bars. The red line is a linear fit through the origin using the first three data points with a slope of 1 emitter / 8.5 kDa, and the dashed line is a constant fit using the last four data points, yielding a constant of 6.5 emitters.*

To demonstrate the applicability of this method to CPs, we investigated the number of independent emitters on a single P3HT chain with respect to molecular weight, i.e. chain length. Single P3HT chains exhibit vastly different spectroscopic properties as a function of their morphology, which is related to their solubility in a particular environment. In general, the solubility of a CP chain decreases as the size of the chain increases, leading to self-folding, which strongly affects excitation energy transfer properties. Therefore, the solubility of CPs affects the size of energy funnels in which efficient singlet-singlet annihilation occurs.[37]

P3HT was fractionated by gel-permeation chromatography (GPC) into six different samples with varying number-average molecular weights, $M_n$, and low polydispersity indices (see Table S1 for details of samples) by comparison to polystyrene standards (see experimental section and ref. [37] for details). Single chains were embedded in a non-polar host matrix, Zeonex, in which most chains unfold,[10] and the samples were encapsulated under a nitrogen atmosphere. The photon statistics were obtained from single P3HT chains by scanning of a 20×20 µm² area (Fig. 1a) at an excitation wavelength of 470 nm, identifying diffraction-limited PL spots and subsequently placing the P3HT chains inside the excitation focus for additional studies. Only the first 250 ms of illumination was considered in evaluating the photon statistics to minimize effects of photo-induced processes (Fig. 1b, red shaded area). From the photon statistics, the median number of emitters per chain was extracted by the CoPS algorithm, and ~200-500 chains per molecular weight sample were measured (see Table S1). Figure 1c shows the dependence of the number of emitters on $M_n$.

A linear increase in the number of emitters could be observed up to a molecular weight of $M_n$ ~ 55 kDa. An average size of ~ 8.5 kDa per emitter was extracted from the slope of the line through the origin (Fig. 1c, red line). Additionally, the number of emitters saturated for $M_n$ > 55 kDa at ~ 6 – 7 emitters (dashed line). The error bars are defined by the lower and upper quartiles of the $N$ distribution and are a measure of the heterogeneity within one molecular weight sample, which is mostly related to polydispersity. The saturation at $M_n$ > 55 kDa is attributed to the self-folding of high-molecular-weight P3HT chains embedded in Zeonex. Strong folding of high-molecular-weight P3HT—achieved, for example, by embedding the chains in a polar environment such as poly(methyl-methacrylate)—can even lead to the existence of a single energy funnel within the polymer chain with corresponding single-photon emission.[10] However, the linear dependence up to $M_n$ ~ 55 kDa suggests that unfolded P3HT chains arise in this size regime, which makes the average size of ~ 8.5 kDa (~ 50 repeat units) per emitter the smallest unit in which efficient singlet-singlet annihilation will always take place in this material.

In addition, we investigated the dynamic changes in the number of emitters, and their average PL brightness, i.e., the detection probability, $p$, was recorded for an extended period lasting up to several seconds. To this end, single

P3HT chains of the $M_n$ ~ 110 kDa sample were placed inside the excitation focus and the PL was recorded. Two examples are shown in Figures 2a and b. Figure 2a displays the overall PL intensity transient (grey curve) together with either the number of independent emitters (top panel, red dots); or with the detection probability (bottom panel, green dots) in steps of 250 ms. The insets are the calculated and normalized cross-correlation curves between the PL intensity and number of emitters transients (top panel, red curve), and between the PL intensity and detection probability transients (bottom panel, green curve). This example shows a strong correlation on the sub-second time scale between the overall PL intensity and the detection probability. The cross-correlation value at zero time delay (0.8) is close to 1, which would correspond to a perfect correlation (bottom panel, inset). No correlation is observed between the PL intensity and the number of emitters (top panel, inset). The example presented in Figure 2b, however, exhibits nearly opposite behavior. No correlation between the PL intensity and detection probability is observed (bottom panel, inset), whereas the PL intensity is clearly correlated with $N$ in this case. The cross-correlation value at zero time delay is 0.6 and decays within the second time regime to zero (top panel, inset). It is important to note that in both cases the decrease in PL intensity is reversible, i.e., blinking is observed, which excludes a simple explanation based on partial bleaching. Additional transients exhibiting the same trends shown in Figures 2a and b are presented in the Supporting Information (Fig. S2a, b), respectively. According to the proposed blinking mechanisms, as illustrated in Scheme 1, the P3HT chains in Figure 2a and S2a contain a shallow charge trap, whereas the P3HT chains in Figure 2b and S2b contain a deep trap. Previous reports on the conjugated polymer poly(2-methoxy-5-(2'-ethylhexyloxy)-1,4-phenylene-vinylene) (MEH-PPV) have made this distinction by only observing the overall PL intensity and sorting blinking events into discrete and gradual changes in intensity.[9, 38] However, the PL intensity transients shown in Figure 2 exhibit both discrete blinking events with almost the same absolute quenching depth of ~ 100 kHz, e.g. after one second in panel a and after twenty-four seconds in panel b. Only a close inspection of the photon statistics can provide the necessary information to distinguish between both quenching events for the material used here. Additionally, in contrast to the PL intensity, the photon statistics provide direct information on the number of independent emitters involved in a quenching process.

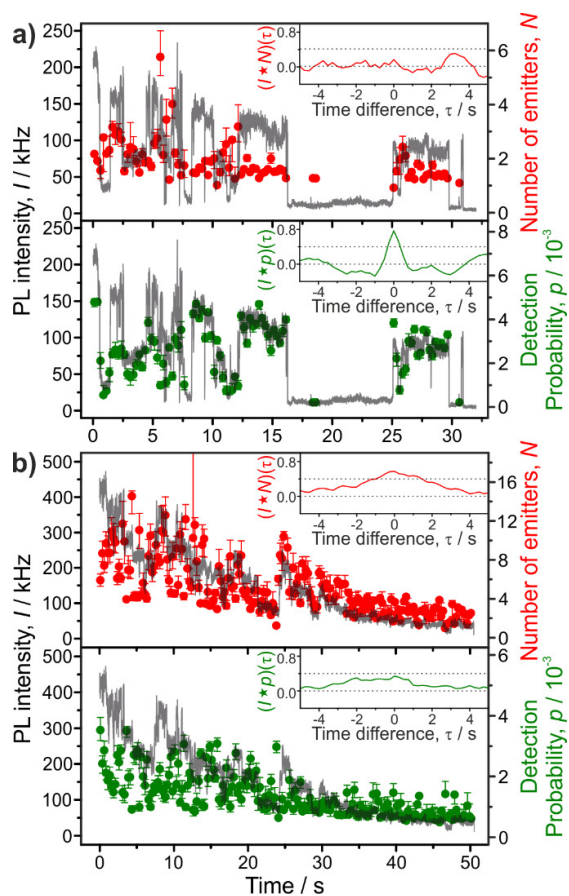

*Figure 2.* PL photon count rate (grey curves) and photon statistics transients are presented for two examples (a and b) of a single P3HT chain embedded in Zeonex. The number of emitters (top panel, red dots) and detection probability (bottom panel, green dots), derived from the CoPS algorithm, are shown including a correlation between the PL intensity, $I$, and the number of emitters, $N$ (inset, red curve), or the detection probability, $p$ (inset, green curve), respectively. The error bars are derived by resampling and repeated fitting of the data (see supporting information for details). Data points, for which the PL intensity dropped below 10 kHz, were omitted, due to an insufficient signal-to-noise ratio.

Finally, the PL intensity transients and photon statistics were recorded from 365 single P3HT chains, which could be sorted according to their blinking behavior into chains containing deep traps (~10 %) or shallow traps (~40 %). The remaining 50 % could not be sorted unambiguously into one of the two blinking types, mainly because the chains exhibit both types of blinking simultaneously or the blinking is partially masked by bleaching. It is conceivable that the formation of a deep or shallow trap is highly morphology dependent, and we hypothesize that a shallow trap is more easily formed in an ordered morphology, whereas a deep trap can be formed solely in a disordered morphology.[39] A highly ordered region can facilitate the mobility of charges,[31, 40] which would expand the quencher´s region of impact, as is typical for a shallow trap (Scheme 1b). In future experiments, excitation polarization measurements will be employed to deduce the morphology of single P3HT chains and correlate their shape with dynamic changes in photon statistics.[41]

In conclusion, we have demonstrated that a detailed investigation of the photon statistics of single CP chains can be used to obtain valuable information about the corresponding blinking mechanism. The recently introduced CoPS method is well suited to study dynamic changes in the photon statistics of CPs at very low photon flux. Two different types of blinking mechanisms are demonstrated for the device-relevant material P3HT, which can be associated with the formation of shallow and deep trap states. A combination of this method with other observables such as the PL lifetime, PL spectrum, and excitation and emission polarization will provide a new lens through which photophysical processes of multi-chromophoric compounds can be studied.

**Experimental Section**

Microscope setup

Single-molecule counting by photon statistics (CoPS) experiments were performed on a custom-built confocal microscope setup with an extended four-detector Hanbury-Brown Twiss detection scheme described elsewhere.[42] The sample was excited by a pulsed picosecond laser diode (PicoQuant, Berlin, Germany) at 470 nm and 20 MHz which was driven by a multichannel diode laser driver (PDL 808 Sepia, PicoQuant, Berlin, Germany). The laser beam was cleaned up by a filter (z473/10) installed in a fiber coupling unit (FCU II, PicoQuant, Berlin, Germany) and collimated after exiting the polarization maintaining fiber (Kollimator MB 02, Linos (now Qioptiq Photonics), Göttingen, Germany) and circularly polarized (achromatic quarter-wave plate, 450 nm to 800 nm, Thorlabs, Munich, Germany) before entering the microscope.

The photoluminescence (PL) of single molecules was collected with a high numerical aperture objective (alpha Plan-Fluar 100x/1.45 oil, immersion oil Immersol 518F $n_e$ = 1.518 (23 °C), both Carl Zeiss, Jena, Germany) and separated from excitation laser light by a dichroic mirror (HC BrightLine BS R488) on a Zeiss Axiovert S100 TV microscope stand (Carl Zeiss, Jena, Germany). The PL passed a telescope array holding a pinhole for confocal detection and a notch filter (single notch filter zet473NF) further blocking scattered excitation light. Four equal beam paths with the PL focused on avalanche photodiodes (APDs, SPCM AQR-14, Perkin-Elmer, Waltham, USA) were created by dividing the PL emission with three 50:50 beam-splitters (G335-520-00, Qioptiq Photonics, Göttingen, Germany). Infrared photons emitted by the APDs during photon detection were excluded by short pass filters (694/SP HC BrightLine, all filters from AHF Analysentechnik, Tübingen, Germany) directly in front of each APD. The detected photons were registered by a SPC 134 system consisting of four synchronized TCSPC-cards SPC 130 (Becker & Hickl, Berlin, Germany) which are controlled by a NI SCB-68 Connector Block (National Instruments, Austin, USA). Scanning and positioning of PL features in the confocal observation volume was performed with a piezostage (piezoelectric scanner P561.3CL with E-503 LVPZT amplifier, E-509.C3A PZT servo-controller for capacitive sensors and E-516.i3 20-bit DAC Interface/Display, Physik Instrumente, Karlsruhe, Germany) with nanometer precision. Stage and photon-counting cards were controlled by custom LabView software (National Instruments, Austin, USA) for synchronized data acquisition.

Preparation and immobilization of poly(3-hexylthiophene)

Single-molecule samples of poly(3-hexylthiophene) were prepared as described in ref. [37]. P3HT (regioregularity: 95.7%, $M_n$: 65.2 kDa, PDI: 2.2; EMD Chemicals Inc., Darmstadt, Germany) was fractionated by gel permeation chromatography (GPC) with a polystyrene standard to obtain 6 samples of different number average molecular weight $M_n$ with low poly dispersion index (PDI) (see Table S1). The P3HT samples were dynamically spin-coated at 2000 rpm from toluene to obtain immobilized and isolated chains of P3HT in a ~200 nm thick Zeonex 480 (Zeon Europe, Düsseldorf, Germany) host-matrix as described below. First, borosilicate glass cover slips (0.17 mm thickness) were cleaned in a 2 % Hellmanex III (Hellma Analytics, Müllheim, Germany) solution and rinsed with water. Then, residual fluorescent molecules were bleached by transferring the glass slides into a UV-ozone cleaner (PSD Pro Series UV, Novascan, Ames, USA). Spin coating on glass cover slides was performed using a mixture of a 0.1-1 pM P3HT/toluene solution and a 6 % Zeonex/toluene solution. Photo-oxidation was prevented by preparing the samples under nitrogen atmosphere in a glovebox and sealing the samples between two cover slips using Araldite®2011 two component epoxy paste adhesive (Huntsman Advanced Materials (Deutschland), Berkamen, Germany).

Acknowledgements

We thank the ERC for financial support through the Starting Grant MolMesON (no. 305070) and the DFG-GRK 1570.

**Keywords:** Single-molecule spectroscopy • Conjugated polymers • Photon statistics • Photophysics

Supporting Information

## Counting by Photon Statistics (CoPS)

CoPS infers the number of fluorescent sites by analyzing the frequency of multiple photon detection events (mDE) in confocal microscopy. In the experiments, the sample is excited by laser pulses shorter than the excited state lifetime with moderate laser repetition rates to ensure relaxation of excited fluorophores between two successive excitation pulses. This scheme guarantees that multiple excitation processes of a single fluorophore can be neglected. The occurrence of mDE thus carries information on the number of emitters in the excitation volume. Figure S1a illustrates the basic counting principle. If only one emitter is present in the confocal detection volume, a laser pulse will lead to emission of either no photon at all, or one photon (blue arrow) in the ideal case without background. If two emitters are present, a single pulse will give rise to the emission of zero, one or two photons at once. The scheme can be generalized for $i$ photons, each detected with a probability of $p$ and corresponding to $N$ individual emitters. Measurement of the probability distribution $P$ of detecting $i$ photons with an efficiency of $p$ allows extraction of the underlying number of emitters $N$.

Typically, a surface scan is carried out to determine the positions of immobilized single probe molecules which are subsequently moved into the laser focus for time-resolved data acquisition. A time correlated single-photon counting (TCSPC) unit with four avalanche photodiodes (APDs) records the photon arrival times for each spot and each APD (for details see Experimental Section) using a modified Hanbury Brown and Twiss detection scheme. The relative occurrence of up to four mDEs are then reconstructed in a histogram, as exemplified in Figure S1b for three different $N$. Background photons that occur in experiments also contribute to the photon statistics and are treated as an additional dim emitter with a low detection probability of $p_b$

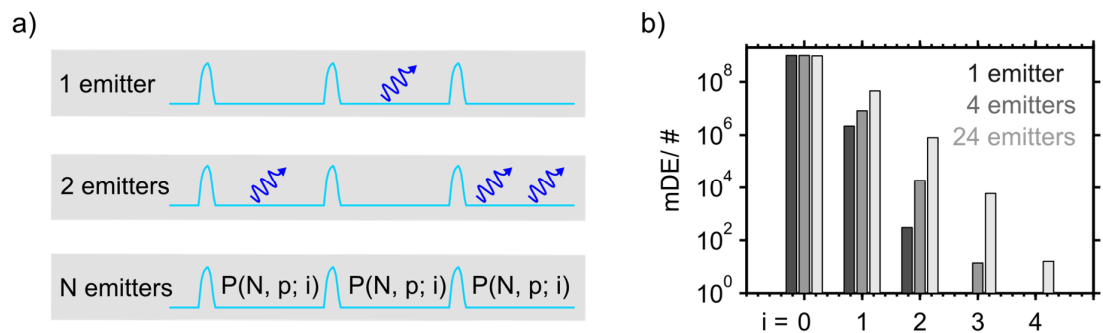

**Figure S1.** Counting by Photon Statistics. a) Scheme describing the occurrence of multiple photon detection events (mDE) after laser excitation. During the experiment, the number of emitters $N$ can vary, so that the experiment can, in principle, be used to extract $N$. b) Expectation values for the relative frequencies for $i$ detection events (i.e. the simultaneous detection of $i$ photons) for 1, 4 and 24 emitters (here: detection probability $p$ = 0.2 %, background detection probability $p_b$ = 0.021 %).

For statistical analysis of the photon stream the full mDE probabilities $P_m(N, p; i)$ are modelled (see Equations S1 and S3-7 and Figure S1b)). They depend on the number of emitters $N$ and the average photon detection probability $p$ of the microscope setup per laser pulse and emitting site (see Equation S2). In the general equations, $m$ denotes the number of detectors (here, $m$=4) while $i$ is the number of photons detected simultaneously. The model accounts for the stochastic processes of excitation, emission and detection of photons including the geometry of the detection path. It should be noted that photon "pile-up" on the detectors is incorporated in the model. That is, when two (or more) photons appear on a single detector after one laser pulse, they will be counted as a single photon due to the dead time of the detector and TCSPC electronics and this is accounted for in the expression given in Equation S1. Background photons in the photon probability distribution are modeled as an additional emitter with low detection probability of $p_b$.[1]

The number of emitters $N$ and their detection probability $p$ is estimated by non-linear regression with a Levenberg-Marquardt algorithm of the model $P_m(N, p; i)$ to the mDE histograms accumulated over the analysis period $t_{acq}$. A resampling method was used to achieve a more robust estimation, i.e. the regression was repeatedly applied to a randomly chosen subset of 75% of all laser cycles in the analysis period $t_{acq}$. After 100 repetitions, we estimate the number of emitters for a single measurement as the median of the fit results and indicate the error by the quantiles $Q_{0.25}$ and $Q_{0.75}$. The background detection probability was estimated by using the CoPS algorithm with $p_b$ = 0 at the end of a trace when the emitters were fully photobleached. The resulting detection probabilities for estimated $N$=1 of ~30 traces were averaged and used as input parameter $p_b$ = 0.021% for the analysis.

## Modeling the Photon Probability Distribution

The photon probability distribution $P_m$ is given by

$$P_m(N, p; i) = \binom{m}{i}\left[\left(1 - \left(\frac{m-i}{m}\right)p\right)^N \left(1 - \left(\frac{m-i}{m}\right)p_b\right) - \sum_{k \geq 0}^{i-1} \frac{\binom{i}{k}}{\binom{m}{k}} P_m(N, p; k)\right] \quad (S1)$$

with $N$ the number of fluorescent emitters, $p$ the photon detection probability (see Eqn. S2), $p_b$ the background photon detection probability, $m$ the number of detectors, $i$ the number of multiple detection events (mDE).

The detection probability $p$ depends on the photon flux, i.e. the average laser intensity $I_{laser}$ divided by the photon energy $h\nu$, the laser repetition frequency $f_{rep}$, the absorption cross-section $\sigma_{abs}$, the fluorescence quantum yield $\Phi_f$ and the overall detection efficiency of the microscope setup $g$.

$$p = \frac{I_{laser}}{f_{rep}h\nu}\sigma_{abs}\Phi_f g = \frac{\varepsilon_{MB}}{f_{rep}} \tag{S2}$$

The molecular brightness $\varepsilon_{MB}$ and the detection probability p are related by the laser repetition frequency. The setup that was used in the experiments had $m=4$ detectors. The photon probability distributions can then be expressed explicitly as:

$$P_4(N,p; i=0) = (1-p)^N(1-p_b) \tag{S3}$$

$$P_4(N,p; i=1) = 4\left(1-\tfrac{3}{4}p\right)^N\left(1-\tfrac{3}{4}p_b\right) - 4(1-p)^N(1-p_b) \tag{S4}$$

$$P_4(N,p; i=2) = 6\left(1-\tfrac{1}{2}p\right)^N\left(1-\tfrac{1}{2}p_b\right) - 12\left(1-\tfrac{3}{4}p\right)^N\left(1-\tfrac{3}{4}p_b\right) + 6(1-p)^N(1-p_b) \tag{S5}$$

$$P_4(N,p; i=3) = 4\left(1-\tfrac{1}{4}p\right)^N\left(1-\tfrac{1}{4}p_b\right) - 12\left(1-\tfrac{1}{2}p\right)^N\left(1-\tfrac{1}{2}p_b\right) + 12\left(1-\tfrac{3}{4}p\right)^N\left(1-\tfrac{3}{4}p_b\right) - 4(1-p)^N(1-p_b) \tag{S6}$$

$$P_4(N,p; i=4) = 1 - 4\left(1-\tfrac{1}{4}p\right)^N\left(1-\tfrac{1}{4}p_b\right) + 6\left(1-\tfrac{1}{2}p\right)^N\left(1-\tfrac{1}{2}p_b\right) - 4\left(1-\tfrac{3}{4}p\right)^N\left(1-\tfrac{3}{4}p_b\right) - (1-p)^N(1-p_b) \tag{S7}$$

## Properties and Measurement Statistics of Fractionated Poly(3-hexylthiophene) Samples

**Table S1.** Number averaged molecular weight $M_n$ of fractionated poly(3-hexylthiophene) samples, Polydispersity index PDI and the respective number of single polymer chains $N$ measured and analysed in single molecule experiments.

|  | 1 | 2 | 3 | 4 | 5 | 6 |
|---|---|---|---|---|---|---|
| $M_n$/ kDa | 19 | 35 | 55 | 71 | 90 | 110 |
| PDI[a] | 1.56 | 1.2 | 1.11 | 1.09 | 1.09 | 1.09 |
| N | 493 | 390 | 238 | 196 | 313 | 365 |

# Additional Example of Correlated PL Intensity Photon Statistics of Poly(3-hexylthiophene) embedded in a Zeonex Matrix

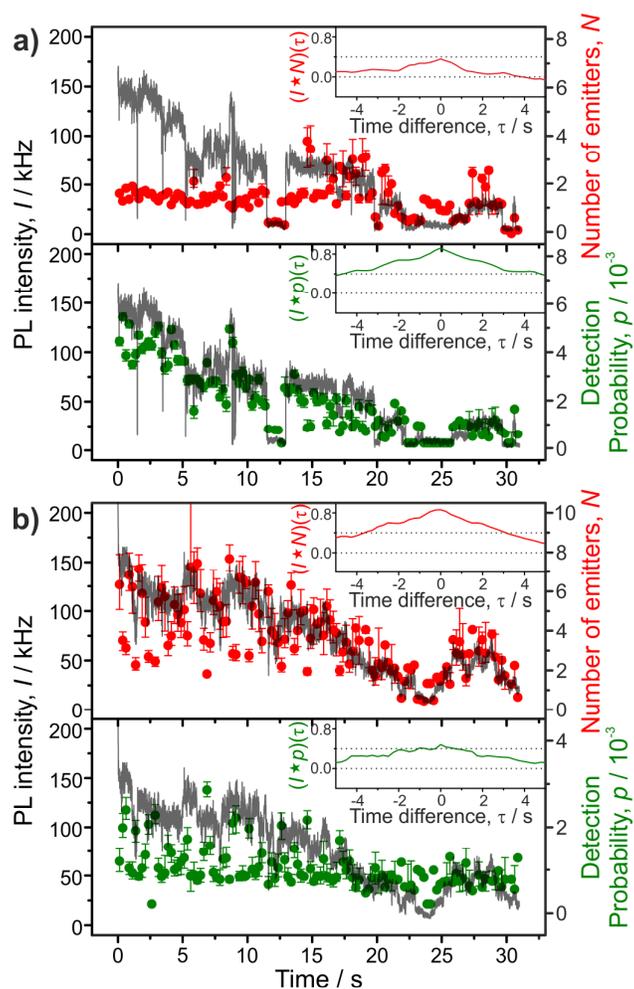

**Figure S2.** PL photon count rate (grey curves) and photon statistics transients are presented for two examples (a and b) of a single P3HT chain embedded in Zeonex. The number of emitters (top panel, red dots) and detection probability (bottom panel, green dots), derived from the CoPS algorithm, are shown including a correlation between the PL intensity, $I$, and the number of emitters, $N$ (inset, red curve), or the detection probability, $p$ (inset, green curve), respectively. The error bars are derived by resampling and repeated fitting of the data.